\documentclass[useAMS,usenatbib,usegraphicx,referee]{mnras}
\usepackage{graphicx,amssymb,amsmath,times,verbatim}
\def\lta{\mathrel{\spose{\lower 3pt\hbox{$\mathchar"218$}}
     \raise 2.0pt\hbox{$\mathchar"13C$}}}
\def\gta{\mathrel{\spose{\lower 3pt\hbox{$\mathchar"218$}}
     \raise 2.0pt\hbox{$\mathchar"13E$}}}

\def\mathnew{\mathsurround=0pt}

\def\simov#1#2{\lower .5pt\vbox{\baselineskip0pt \lineskip-.5pt
\ialign{$\mathnew#1\hfil##\hfil$\crcr#2\crcr\sim\crcr}}}

\title[]{Multi-wavelength Temporal Variability of the Blazar 3C 454.3 during 2014 Activity Phase}
\author[P. Kushwaha et al.]{Pankaj Kushwaha$^{1, 2}$\thanks{E-mail:pankaj.kushwaha@iag.usp.br},
Alok C. Gupta$^{3,4}$\thanks{CAS PIFI Fellow}, Ranjeev Misra$^{1}$, K. P. Singh$^{5}$\\
$^{1}$ Inter-University Center for Astronomy \& Astrophysics (IUCAA), Pune 411007, India \\
$^{2}$ Present address: Department of Astronomy (IAG-USP), University of Sao
Paulo, Sao Paulo 05508-900, Brazil \\
$^{3}$ Key Laboratory for Research in Galaxies and Cosmology, Shanghai Astronomical
Observatory, Chinese Academy of Sciences, \\
~~~80 Nandan Road, Shanghai 200030, China \\
$^{4}$ Aryabhatta Research Institute of Observational Sciences (ARIES), Manora Peak,
Nainital 263002, India \\
$^{5}$ Department of Astronomy \& Astrophysics, Tata Institute of Fundamental Research
(TIFR), Mumbai 400005, India
}

\begin{document}
%  \linenumbers

\maketitle

\begin{abstract}
We present a multi-wavelength temporal analysis of the blazar 3C 454.3 during 
the high $\gamma$-ray active period from May-December, 2014. Except for
X-rays, the period is well sampled at near-infrared (NIR)-optical by the \emph{SMARTS}
facility and the source is detected continuously on daily timescale in the \emph{Fermi}-LAT 
$\gamma$-ray band. The source exhibits diverse levels of variability with many flaring/active
states in the continuously sampled $\gamma$-ray light curve which are also reflected
in the NIR-optical light curves and the sparsely sampled X-ray light curve by the
\emph{Swift}-XRT. Multi-band correlation analysis of this continuous segment during
different activity periods shows a change of state from no lags between IR and $\gamma$-ray,
optical and $\gamma$-ray, and IR and optical to a state where $\gamma$-ray
lags the IR/optical by $\sim$3 days. The results are consistent with the previous
studies of the same during various $\gamma$-ray flaring and active episodes
of the source. This consistency, in turn, suggests an extended localized emission
region with almost similar conditions during various $\gamma$-ray activity states.
On the other hand, the delay of $\gamma$-ray with respect to IR/optical and a trend
similar to IR/optical in X-rays along with strong broadband correlations favor
magnetic field related origin with X-ray and $\gamma$-ray being inverse Comptonized of 
IR/optical photons and external radiation field, respectively.
\end{abstract}


\section{Introduction} \label{sec:intro}
Blazars are jetted active galactic nuclei (AGNs) with relativistic jets align
at close angles to observer's line of sight. They are characterized by a highly
variable, predominantly non-thermal
continuum emission spanning the entire accessible electromagnetic spectrum
with a significant polarization at radio-to-optical wavelengths, and
superluminal features in high-resolution radio imaging \citep{2013AJ....146..120L}.
In the temporal domain, flux variability is seen on all timescales ranging from
minutes to years and is believed to be a manifestation of Doppler boosting
associated with the close alignment of the relativistic jet with the line of sight.
Traditionally, blazars have been classified as BL Lacertae objects (BL Lacs) and
flat spectrum radio quasars (FSRQs) based on the absence and presence of prominent
broad emission lines in their optical-ultraviolet spectra \citep{1995PASP..107..803U}.

Despite a wide range of variability in energy and time domains, blazars spectral
energy distributions (SEDs) exhibit a characteristic broad double-hump profile
\citep{1998MNRAS.299..433F, 2016arXiv160403856M}. The low energy hump peaks between
infra-red (IR) to ultraviolet(UV)/X-rays, and is widely accepted to be due to synchrotron
emission from relativistic non-thermal electrons in the jet. The emission at high
energy hump which peaks at $\gamma$-ray energies is still unclear and, is well
reproduced by both leptonic and/or hadronic non-thermal processes \citep[e.g.][]
{2013ApJ...768...54B, 2015MmSAI..86...13D}. In the leptonic models, the high energy emission
originates as a result of inverse Compton (IC) scattering \citep[e.g.][]{2014Natur.515..376G} of 
ambient photons which can be synchrotron photons and/or photons external to the jet,
like photons from the broad-line region (BLR), torus photons and/or Cosmic Microwave
Background (CMB) photons. The hadronic models, on the other hand, attribute it
to the interactions of relativistic protons in the jet with the magnetic field
\citep[proton synchrotron,][]{2001APh....15..121M} and/or with the soft radiation field
\citep[photo-pion cascade,][]{1992A&A...253L..21M}. 

Understanding the nature of variability in Blazars has eluded the researchers
over the years. Generally attributed to relativistic shocks and/or magnetic reconnection
processes, it  differs from source-to-source, and even during different activity states of
a source. Furthermore, its highly energy dependent manifestation across the electromagnetic
(EM) spectrum makes it complex to decipher. This makes multi-wavelength spectral
and temporal study of blazars emission one of the potential tools to probe and
understand the physical conditions/processes responsible for its energy dependent
variability within the compact unresolvable sites \citep[e.g.][]{2014ApJ...796...61K}.
The correlations between different wavelengths carry an imprint of dynamics
of interplay between energization and losses and hence, their relative dominance
in different energy bands.

3C 454.3, located at the redshift of $\rm z = 0.859$ is a bright and a highly
variable FSRQ first detected at $\gamma$-ray energies ($> 100$ MeV) by the \emph{EGRET}
telescope onboard the CGRO \citep{1993ApJ...407L..41H}. The source has been studied
extensively at different wavelengths over the last two decades. However, only after 
2005 activity which was seen in all the accessible window of the electromagnetic
spectrum \citep{2006A&A...453..817V, 2006A&A...456..911G, 2006A&A...449L..21P}, that it
became one of the targets of coordinated multi-wavelength studies.

3C 454.3 has been extremely active FSRQ at $\gamma$-ray energies since 2007 as seen
by \emph{AGILE} \citep{2010ApJ...718..455S} as well as the scanning $\gamma$-ray
observatory \emph{Fermi}-LAT post its launch in 2008. Many extraordinary $\gamma$-ray
activities in terms of spectral and temporal variations \citep{2010ApJ...721.1383A,
2011ApJ...733L..26A, 2015arXiv151102280B} have been reported with counterparts in other
parts of the electromagnetic spectrum. The coordinated follow-ups 
during many of these high $\gamma$-ray activity periods have
revealed diversity and complexity of emission processes in the source.
The \emph{AGILE} 2007 multi-wavelength campaign observed a correlated optical and
$\gamma$-ray variation with no lags during the November \citep{2009ApJ...690.1018V},
but found a possible $\lesssim 1$ day lag with $\gamma$-ray lagging the optical
during the December observations \citep{2009ApJ...707.1115D}. A more extensive
multi-wavelength campaign led by \emph{AGILE} over 18 months found almost
simultaneous peaks in different energy bands with a delay of less than a day
\citep{2010ApJ...712..405V}, consistent with its previous finding. Similarly, the
multi-wavelength observation during \emph{Fermi}-LAT operation by \citet{2009ApJ...697L..81B}
for the period of August to December 2008 revealed an excellent correlation between the IR, optical,
UV, and gamma-ray light curves with a time lag of less than one day but no correlation
between X-ray flux with either of these EM bands. A similar result was found
for another multi-wavelength data set compiled by \citet{2012ApJ...756...13B} for 
the period of June 2008 to December 2010 showing excellent correlations between the IR,
optical, and $\gamma$-rays with a time lag of less than a day, while 
\citet{2012ApJ...758...72W} have found near-simultaneous
variations in millimeter, far-IR and $\gamma$-rays with $\gamma$-ray lagging
IR (160 $\rm \mu m$) by $1\pm0.5$ days for November 2010 - January 2011 period. 
On the other hand, a study by \citet{2012AJ....143...23G} for November-
December 2009, has found a lag of $\sim 4.5$ days with $\gamma$-ray leading
optical, but neither being correlated with X-rays. Thus, the broadband emission
during various $\gamma$-ray activity states seems to have recurring features 
despite widely different variability amplitudes in both flux and time domains.
The well sampled simultaneous/contemporaneous data set generated by the coordinated
follow-ups across the EM spectrum by the ground and space based observatories
in response to the \emph{Fermi}-LAT triggers have, thus, opened a window
for systematic exploration of various characteristics associated with
particular sources \citep[e.g.][]{2016ApJ...822L..13K}, thereby providing insights
and constraints on the rich physics of the relativistic jets, emission region etc.

Here, we present correlation analysis of multi-wavelength data during 
a $\gamma$-ray active period between May-December, 2014, when the source 
$\gamma$-ray fluxs over daily timescale were $\rm\gtrsim 10^{-6}~ ph~ cm^{-2}~ s^{-1}$
and was followed in other electromagnetic bands. The source exhibited high $\gamma$-ray variability
of different levels which were also noticed in X-rays and NIR-optical bands.
The paper is organized into five sections with the next section presenting the details
of data resources and associated reduction processes. Section 3 presents the temporal
analysis technique and results, followed by discussion and implications in Section 4.
We finally conclude in Section 5. 

\section{Multi-wavelength Archival Data and Reduction} \label{sec:data}
We have made use of publicly available multi-wavelength data from $\gamma$-rays
to NIR-optical. The $\gamma$-ray and X-ray data are taken from their respective
data archive centers and reduced following the procedures recommended by the respective
instrument teams. The corresponding contemporaneous/simultaneous NIR-optical data are
taken from the \emph{SMARTS} facility supporting the \emph{Fermi Multiwavelength AGN
Science}. The multi-wavelength observations preseneted here are part of
coordinated follow-ups when the source flux in \emph{Fermi} Large Area Telescope (LAT)
approaches $\rm 10^{-6} ph~cm^{-2}~s^{-1}$ \citep{2012IAUS..285..294C}.

\subsection{Fermi $\gamma$-ray Data}
The LAT onboard \emph{Fermi Gamma-ray Space Telescope} is a
pair conversion imaging telescope which normally operates in scanning mode. It is sensitive
to $\gamma$-ray photons with energy $>$ 20 MeV \citep{2009ApJ...697.1071A} and  covers
the entire sky every $\sim$3 hours. Here, we have used the LAT data of
3C 454.3 from May 13th, 2014 to December 24, 2014 (MJD: 56790-57015) and analyzed using
the \emph{Fermi Science Tool} version v10r0p5 with appropriate selections and cuts
recommended for the scientific analysis of \emph{PASS8} data. 

The photon like events categorized as ``evclass=128, evtype=3'' with energies
0.1$\leq$E$\leq$300 GeV within a maximum zenith angle of 90$^\circ$ were selected
from a circular region of interest (ROI) of $15^\circ$ centered on the source. The
appropriate good time intervals were then generated by using the recommended criteria 
``(DATA\_QUAL$>$0)\&\&(LAT\_CONFIG==1)''. The likely effects of cuts and selections,
as well as the presence of other sources in the ROI, were incorporated by generating
exposure map on the ROI and an additional annulus of $10^\circ$ around it. These
events were then modeled using `unbinned likelihood analysis' with input model file
from the 3rd LAT catalog \citep[3FGL --gll\_psc\_v16.fit;][]{2015ApJS..218...23A}\footnote{
3C 454.3 is fitted with a log parabola model}. The Galactic and isotropic 
extragalactic contributions were accounted by using the respective templates,
\emph{gll\_iem\_v06.fits} and \emph{iso\_P8R2\_SOURCE\_V6\_v06.txt} file provided by
the instrument team. A significance criterion of $3\sigma$ corresponding to a TS
(Test Statistic) value of $\sim10$ has been used for the source detection.

\subsection{Swift X-ray Data}
The \emph{Swift}-XRT data during the mentioned period were analyzed using the methods
suggested by the instrument team with the default parameters setting. Only pointed
observations from the window timing (WT) or photon counting (PC) modes have been
used. Each XRT observation is processed with the \emph{xrtpipeline} task using the
latest \emph{CALDB} files within \emph{heasoft-6.18}. Higher order products like
source light-curves and spectra were generated with \emph{xselect} using a source
region of $\rm 47.2''$ \citep[90\% PSF,][]{2005SPIE.5898..360M}. Most of the
PC mode data have count rates above the pile-up free limit of 0.5 counts/s. Such
observations were accounted for the pile-up using an annulus region, leaving the
central 4-8 pixel radius corresponding to the departure of the XRT PSF model
(King\footnote{$\rm PSF(r) = [1+(r/r_c)^2]^{-\beta}; r_c = 5.8, \beta = 1.55$}
profile) and the data \citep{2006ApJ...638..920V}\footnote{http://www.swift.ac.uk/analysis/xrt/pileup.php}.
The departure point was estimated by first fitting the King profile to the outer
wing ($\rm \gtrsim 15~pixels$) of the PSF with normalization being the only free parameter
and then extrapolating this best fit to the center.
Accounts of instrument related effects and various selections on data are compensated
by generating ancillary response file through \emph{xrtmkarf} task. The resulting
spectra were then modeled using \emph{phabs*powerlaw} within \emph{XSPEC} with a
fixed neutral hydrogen column density (nH) of $1.34\times 10^{21} ~\rm cm^{-2}$
as deduced from Chandra observations\footnote{XRT observations with good exposure
($\gtrsim$ 5000 s) also suggest similar nH values.} of 3C 454.3 \citep{2006A&A...453..817V}.
The best model parameters were then used to calculate the unabsorbed 0.3-10.0 keV
flux using \emph{cflux}.

\subsection{NIR-Optical Data from \emph{SMARTS}}
The near-infrared (NIR)-optical photometric data are taken from the 
\emph{SMARTS}\footnote{http://www.astro.yale.edu/smarts/glast/home.php}
blazar monitoring campaign, supporting the \emph{Fermi Multiwavelength AGN Science}.
Except for reddening corrections, the data are publicly available for Science use.
The details of instruments and data reduction processes are mentioned in \citet{2012ApJ...756...13B}. 
Here, we have used all the available data during the high $\gamma$-ray activity
period, corrected for reddening using an E(B-V) value of $0.0889 \pm 0.0041$ following
\citet{2011ApJ...737..103S}.

\begin{figure}%[h]
\begin{center}
 \includegraphics[scale=1]{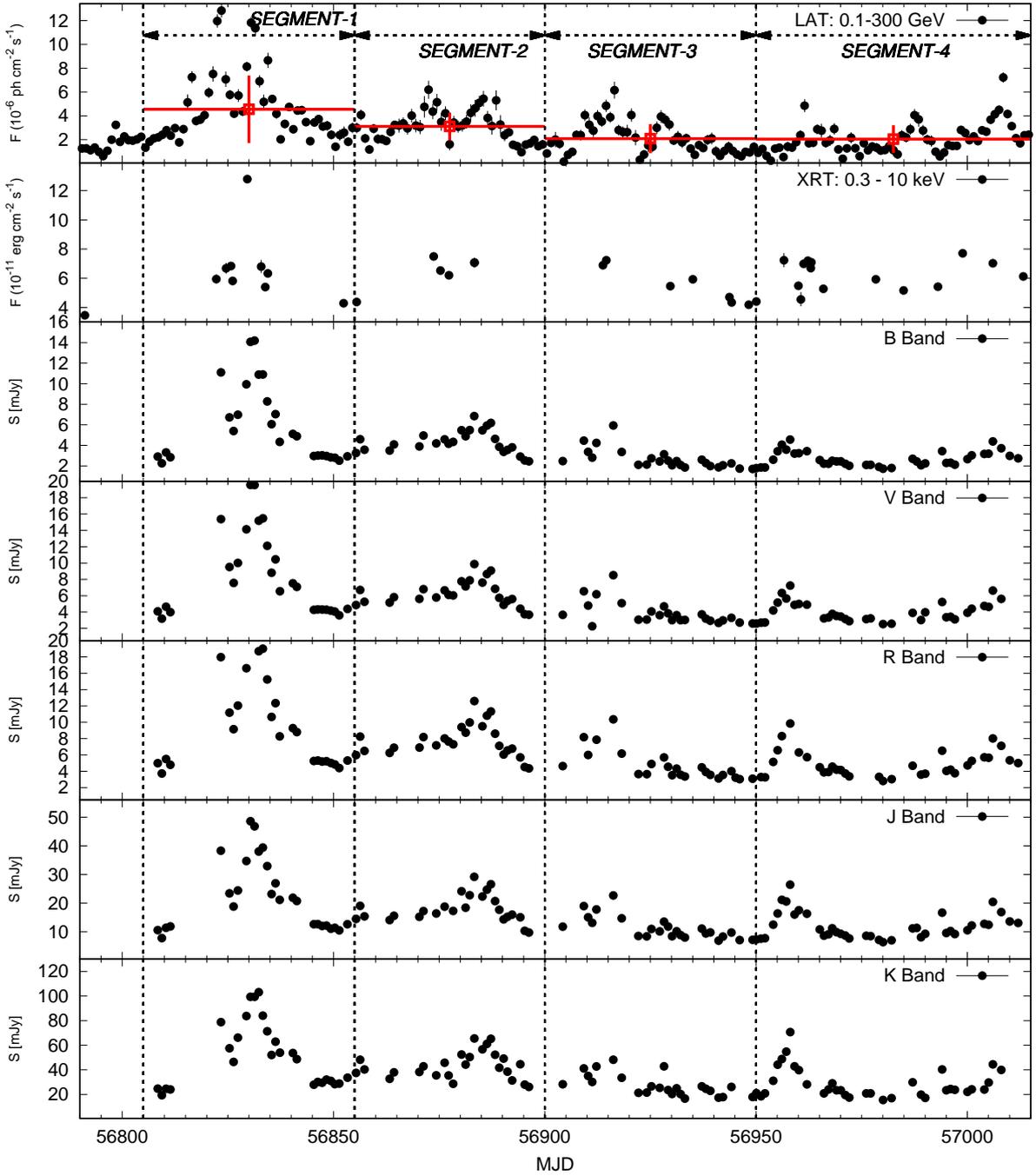}
\end{center}
\caption{Multi-wavelength light curves of 3C 454.3 from $\gamma$-rays to IR-Optical
during a high $\gamma$-ray activity period from May 13 - December 24, 2014.
The LAT light curve is extracted on daily timescale with X-ray from the publicly
available data and NIR-optical from \emph{SMARTS} follow up (see \S\ref{sec:data}).
The vertical lines demarcate the sections considered for temporal analyses
in the present work while the red point represent the respective mean with intrinsic variance
as its error (see \S\ref{sec:analysis}).}
\label{fig:mwlc}
\end{figure}

\subsection{Multi-wavelength Light Curves}\label{sec:analysis}
The multi-wavelength light curves from $\gamma$-rays to IR-optical between May 13th
to December 24th, 2014 are shown in Figure \ref{fig:mwlc}. Varied levels of
variability in both flux and time from IR-optical to $\gamma$-ray energies at different
times can be seen. Though X-ray data are not well sampled, the trend does reflect
the activity seen in the IR-optical and $\gamma$-ray light curves. Thus, based on
 $\gamma$-ray variability and sampling of data at other energies (IR-optical),
the multi-wavelength light curve is divided into four segments; SEGMENT-1: MJD
56805-56855, SEGMENT-2: MJD 56855-56900, SEGMENT-3: MJD 56900-56950 and SEGMENT-4:
MJD 56950-57015 demarcated by the vertical lines in Figure \ref{fig:mwlc} to
further investigate the variability. In addition, we have plotted
the segment wise mean with intrinsic variance as its error for the $\gamma$-ray light
curve (red points, Fig. \ref{fig:mwlc}) with fluxes for the entire duration being 
consistent with a log-normally distribution \citep[e.g.][]{2016ApJ...822L..13K}. 
Flares in the $\gamma$-ray band, on the other hand, are defined
using the segment's variance: fluxes inconsistent with the segment's intrinsic 
variance and total duration (rise + fall) of $\lesssim$ a week (termed
as \emph{strong flares} henceforth).

\section{Variability Analysis and Results}
We have carried out cross-correlation analysis between $\gamma-$ray vs optical,
$\gamma-$ray vs NIR, optical vs optical, NIR vs NIR, and optical vs NIR using the
\emph{z-transformed discrete correlation function} (ZDCF) method \citep{2013arXiv1302.1508A,
1997ASSL..218..163A}. The method uses equal population binning and Fisher's z-transform
to account for biases introduced by sparse, uneven sampling of data. It estimates
the correlation coefficients using the data pairs sorted according to their lags
with at least 11 (default) pairs in a bin, after the removal of interdependent
data pairs from each bin. The errors on the coefficients, on the other hand,
are estimated by employing Monte Carlo simulation of the light curve by taking into
account the observational errors on the fluxes. For each instance of simulated light
curve pair, the estimated correlation coefficients are transformed to the z-space via 
\citep{2013arXiv1302.1508A, 1997ASSL..218..163A}
\begin{equation}
 z = \frac{1}{2} log\left(\frac{1+r}{1-r}\right), \qquad 
 \zeta = \frac{1}{2} log\left(\frac{1+\rho}{1-\rho}\right), \qquad
 r = tanh(z) \nonumber
 \end{equation}
where $\rm r$ is the correlation coefficient and $\rm \rho$ is the unknown population
correlation coefficient of the bin. The transformed quantities are normally distributed
and the ZDCF uses an ansatz $\rm \rho = r$ to estimate the mean and variance of the
$z$ \citep[for more details]{2013arXiv1302.1508A}. Theses errors are then transformed
to the correlation space providing $1\sigma$ errors on the correlation coefficients.

In the present study, we generated 1000 realizations of each light curve pair to
estimate the error on the DCF values. The resulting DCF plots between different light 
curves for all the four segments are shown in Figure \ref{fig:zdcf} and the corresponding
lag values are reported in Table~\ref{tab:lagResutls}. A positive lag for
DCF labeled as ``LC1 vs LC2'' means LC2 emission lagging LC1 and vice-versa.

% \subsection{Correlated Variability}
In the cross-correlation analysis, we have not used X-ray observations taken with Swift/XRT 
in 0.5 -- 10 keV energies and plotted in the 2nd panel from top in Figure 1. The X-ray
data, though sparsely sampled are very few for the ZDCF analysis which by default requires
at least 11 data pairs to form a bin. Data in optical and NIR bands, on the other hand,
have almost identical sampling and on visual inspection appear very well 
correlated without any lag. To avoid many plots, we used V and R optical bands LCs for 
correlation with $\gamma-$ray, and optical-optical correlation. NIR J band LCs are used 
for correlation with $\gamma-$ray, NIR K band, and optical R band LCs. Cross-correlated 
plots for four segments are plotted in Figure \ref{fig:zdcf} and correlation done between
different bands are marked in different panels.     

\subsection{SEGMENT 1}
\emph{SEGMENT 1} is taken from MJD 56805 to 56855 which shows three \emph{strong 
flares} in $\gamma$-ray peaked at MJDs: 56823.5, 56830.5 and 56834.5, and appear nearly
simultaneous (within observation cadence) in optical, NIR, and X-rays as well (see Figure
\ref{fig:mwlc}). The ZDCFs
are plotted in the 1st column from left of Figure \ref{fig:zdcf}, and cross-correlation
results are reported in the 2nd column of Table \ref{tab:lagResutls}. The results
in Figure \ref{fig:zdcf} and Table \ref{tab:lagResutls} show the strong flare detected
in all EM bands are simultaneous (i.e. ZDCF lag is 0 within error). The flare emission
is thus, likely co-spatial and hence, intrinsic to the source.   
 
\subsection{SEGMENT 2}
\emph{SEGMENT 2} is taken from MJD 56855 to 56900 which shows three
\emph{strong flares} in LAT peaked at MJDs: 56872.5, 56885.5, and 56888.5
(see trends in X-rays) but only two similar trends in optical and NIR bands (see
Figure \ref{fig:mwlc}). The trends in optical and NIR  bands visually appear
simultaneously with 2nd flare in $\gamma-$ray. The corresponding ZDCFs are
plotted in the 2nd column from left of Figure \ref{fig:zdcf}, and cross-correlation
results are reported in the 3rd column of Table \ref{tab:lagResutls}. The results
in Figure \ref{fig:zdcf} and Table \ref{tab:lagResutls} show the 2nd $\gamma-$ray
flare detected in all EM bands are simultaneous (i.e. ZDCF lag is 0 within error).
The 1st gamma-ray flare is not observed in optical and NIR bands. Thus, similar to
\emph{SEGMENT 1}, 2nd $\gamma-$ray flare emission is intrinsic in nature with the
same emitting region in optical and NIR bands.  

\subsection{SEGMENT 3}
\emph{SEGMENT 3} corresponds to MJD 56900 to 56950 and exhibits four \emph{strong
flares} in $\gamma-$ray peaked at MJDs: 56909.5, 56916.5, 56920.5, and 56927.5 with
the second being the most prominent. Further, the first two and the
last $\gamma-$ray flares are also accompanied in the optical and NIR bands, though
the data sampling is comparatively poor (see Figure \ref{fig:mwlc}). The corresponding
ZDCFs are plotted in the 3rd column from left in Figure \ref{fig:zdcf} with results
reported in the 4th column of Table \ref{tab:lagResutls}. Though the DCFs are flatter
and chaotic especially the optical-optical/NIR correlations, the results in Figure
\ref{fig:zdcf} and Table
\ref{tab:lagResutls}, nevertheless, show that the first prominent flare is detected
nearly simultaneous in $\gamma-$ray, optical, and NIR bands (i.e. ZDCF lag is 0 within
error). The flare emission is therefore, just like the flares during the \emph{SEGMENTs}
1 and 2.
 
\subsection{SEGMENT 4}
\begin{figure}%[!ht]
\centering
 \includegraphics[scale=0.65]{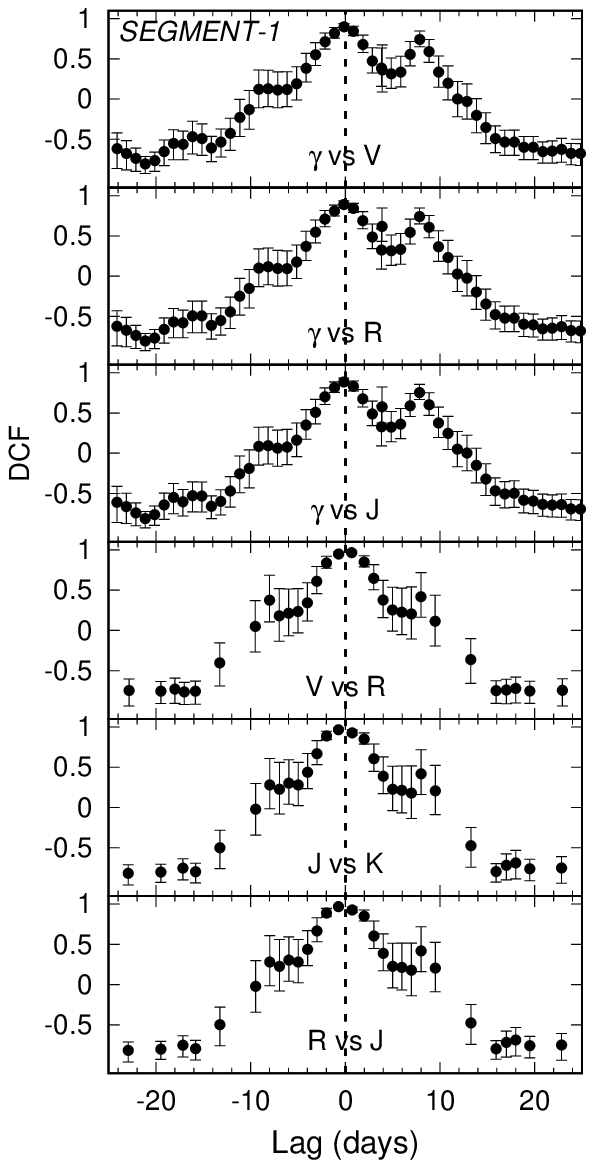}
 \includegraphics[scale=0.65]{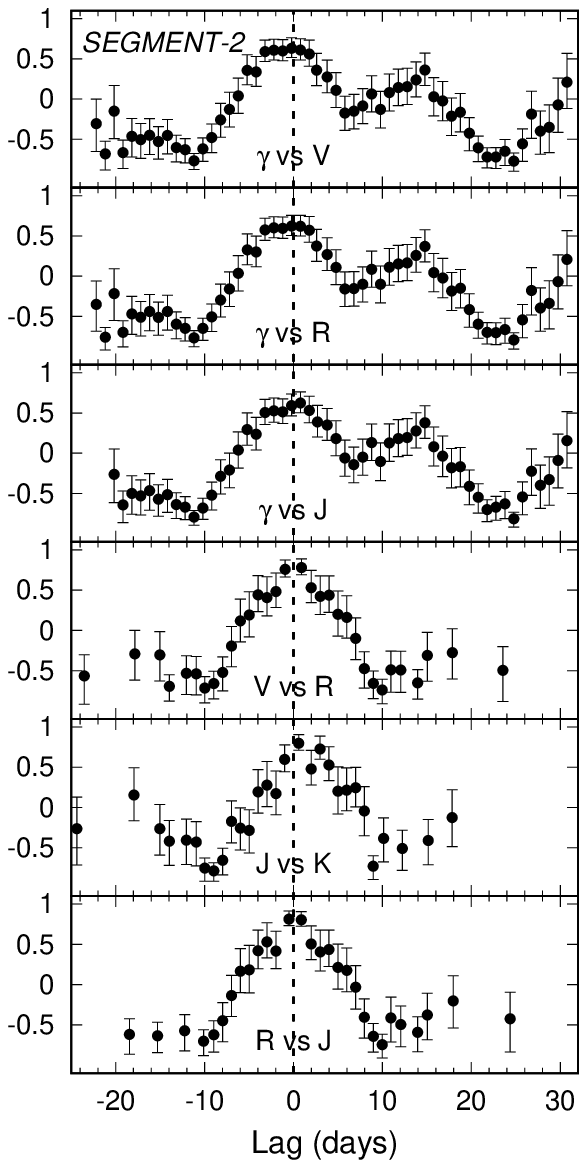}
 \includegraphics[scale=0.65]{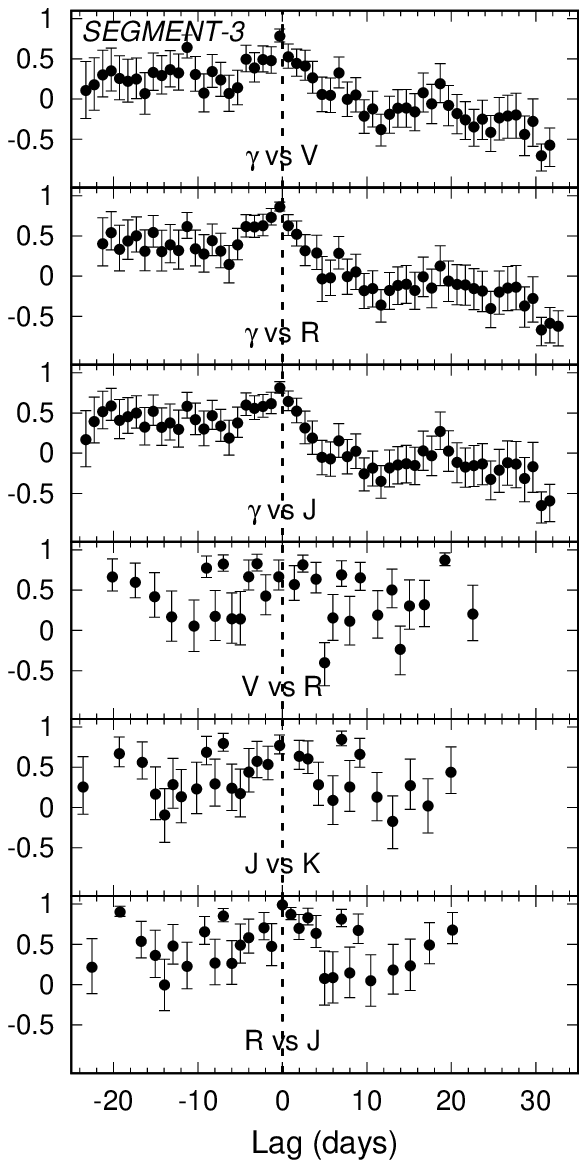}
 \includegraphics[scale=0.65]{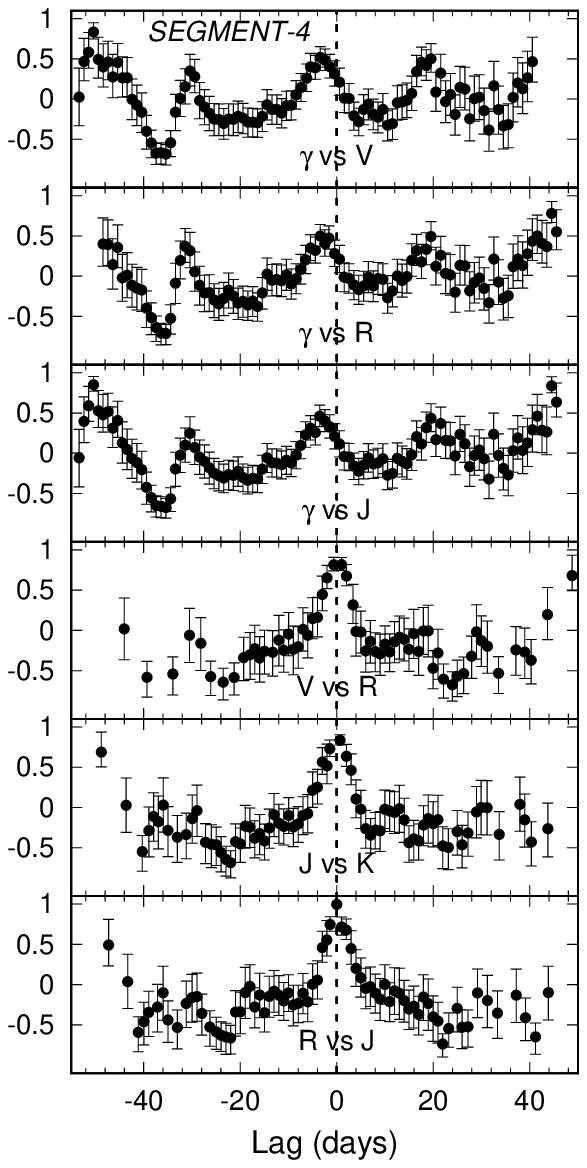}
 \caption{DCF between $\gamma$-optical, $\gamma$-IR and optical-IR light curves for
 the 4 different segments of the near continuous multi-wavelength light curves of
 3C 454.3 (see Figure \ref{fig:mwlc}). The vertical lines correspond
 to zero lag between the light curves labeled as ``LC1 vs LC2''.}
 \label{fig:zdcf}
\end{figure}

\emph{SEGMENT 4} is taken from MJD 56950 to 57015 and exhibits three \emph{strong
flares} in $\gamma-$ray peaked at MJDs: 56961.5, 56987.5 and 57008.5. Similar trends can also
be seen in the well sampled optical and NIR bands, though a few days earlier ($\sim$ 3
days, see Figure \ref{fig:mwlc}). The {ZDCFs} are plotted in the 4th column from
left of Figure \ref{fig:zdcf} and show many (suggest two) peaks of similar
strengths for $\gamma$-ray-optical/NIR (optical-optical/NIR) correlations. However,
we select the DCF peak closest to the zero lag as others (except one at +20) are
almost at the edge of the plot where the lags are of order of the data length used for
estimating the correlations and are likely due to the three strong $\gamma$-ray
flares in the segment correlating with the strongest optical/NIR fluxes in the
segments (also supported by their separation in the light curves, see Fig. \ref{fig:mwlc}). 
The cross-correlation results corresponding to this DCF peak are reported in the
5th column of Table \ref{tab:lagResutls}. The results in Figure \ref{fig:zdcf} and
Table \ref{tab:lagResutls} show the prominent rise/fall detected in $\gamma-$ray
lag by $\sim$3 days with respect to similar features seen in the optical and NIR
bands. This is also supported by the well sampled optical-NIR data around the first and the
last $\gamma$-ray flare. Thus, apart from flare emission being intrinsic, it hints that
the emission probably originates from a different emitting regions for the optical/NIR
and $\gamma-$ray bands and/or an act of magnetic field within the same region (see
\S\ref{sec:discussion}).

\begin{table}\label{tab:lagResutls}
  \centering
% \begin{minipage}{140mm}
  \caption{Lag results for all the segments (in days)}
  \label{tab:lagResutls}
  \begin{tabular}{ccccc}
  \hline
Light curves & \emph{SEGMENT-1} 	& \emph{SEGMENT-2} 	  & \emph{SEGMENT-3} 	    & \emph{SEGMENT-4} \\ \hline
$\gamma$ vs V & $-0.14^{+0.68}_{-0.59}$ & $-0.27^{+1.51}_{-2.3}$  & $-0.31^{+0.54}_{-0.58}$ & $-3.43^{+1.43}_{-1.53}$ \\ 
$\gamma$ vs R & $-0.15^{+0.78}_{-0.61}$ & $-0.21^{+1.55}_{-2.22}$ & $-0.31^{+0.47}_{-0.67}$ & $-3.43^{+1.85}_{-1.38}$ \\
$\gamma$ vs J & $-0.15^{+0.74}_{-0.66}$ & $+0.78^{+1.06}_{-2.39}$ & $-0.31^{+0.56}_{-0.73}$ & $-3.51^{+1.38}_{-1.56}$ \\
V vs R & $+0.66^{+0.59}_{-1.08}$ & $+0.89^{+0.69}_{-1.64}$ & $-3.0^{+5.6}_{-0.6}$    & $+1.0^{+0.6}_{-1.7}$ \\
R vs J & $-0.72^{+0.83}_{-0.59}$ & $-0.51^{+1.36}_{-0.78}$ & $-0.01^{+0.44}_{-0.56}$ & $-0.01^{+0.44}_{-0.60}$ \\
J vs K & $-0.72^{+0.83}_{-0.59}$ & $+0.57^{+2.21}_{-0.86}$ & $-0.3^{+1.9}_{-1.7}$    & $+0.68^{+0.66}_{-1.25}$ \\
  \hline
  \end{tabular}
%  \end{minipage}
% \label{tab:lagResutls}
 \end{table}

\section{Discussion}\label{sec:discussion}
The correlation analyses performed on a few continuous segments of multi-wavelength
light curves exhibit a wide variety of features, ranging from nicely peaked to flat
DCF between different energy bands. These DCF profiles signify the inter-relation
and relative contributions of various emission components as reflected in the
multi-wavelength light curves. Since the appearance of the biggest $\gamma$-ray
flare (\emph{SEGMENT 1}) to the last segment (\emph{SEGMENT 4}), the lags between
IR/optical-$\gamma$-rays change from zero to a lag of $\sim$ 3 days with
IR-optical leading $\gamma$-rays. On the other hand, the IR-optical remains consistent
with no lags throughout (see Figure \ref{fig:zdcf}, and Table \ref{tab:lagResutls}).
The profiles too become more or less similar  in all the bands by the end i.e.
\emph{SEGMENT} 4. The correlation analyses during \emph{SEGMENT 1} show a double
peak in IR/optical-$\gamma$-rays but a single peak in IR-optical correlations. The
highest value DCF peak, however, is consistent with zero lag for all the bands (see
Table \ref{tab:lagResutls}). \emph{SEGMENT 2} also shows a similar nature and results.
The correlation for \emph{SEGMENT 3} are comparatively flatter across all the bands
compared to the rest (SEGMENTS) with the peak values being consistent with zero lags.
\emph{SEGMENT 4}, contrary to all, exhibits a lag of $\sim$ 3 days for 
$\gamma$-ray with respect to IR-Optical.

The highly correlated and nearly simultaneous variations across the electromagnetic
spectrum strongly indicate a co-spatial origin with the same population powering
the emission at all energies, at least during the flares. This, in turn, favors
leptonic processes where the non-thermal relativistic electrons emitting synchrotron
emission scatter the low energy seed photons to higher energies via IC.
In 3C 454.3,
multi-wavelength correlation studies in the past and the spectral analysis of the 
$\gamma$-ray flares suggest the emission to be mainly originating at the boundary
of the BLR and/or beyond it i.e. at the torus scales \citep{2014MNRAS.441.1899F,
2016arXiv160107180C, 2015arXiv151102280B}. For a given lepton population $\rm N(\gamma)$,
with $\rm \gamma$ being lepton Lorentz factor, the observed radiation
due to synchrotron and IC of external seed photons follow the dependency
\citep{2013MNRAS.433.2380K, 2012MNRAS.419.1660S}
\begin{align}
 F_{syn}(\nu) &\propto B^2 \nu_L^{-3/2} N\left(\sqrt{\frac{\nu}{\nu_L}}\right)
 \nu^{1/2} \nonumber \\
 F_{IC}(\nu) &\propto \frac{u_{iso, \ast}}{\nu_\ast} \sqrt{\frac{\Gamma (1+\mu) \nu}{\nu_\ast}}
 N\left(\sqrt{\frac{\nu}{\Gamma (1+\mu) \nu_\ast}}\right) \nonumber
\end{align}
where $\rm \Gamma$ is the bulk Lorentz factor, B is the emission region magnetic
field, and $\rm \mu=cos\theta$ with $\rm \theta$ being the angle with respect to our line 
of sight. $\rm u_{iso, \ast}$ and $\rm \nu_\ast$ refers to the energy density and peak
frequency of the isotropic thermal photon field external to the jet in the AGN frame while
$\rm \nu_L = eB/2\pi m_e c$ is the Larmor frequency of electron of mass $m_e$, and c is the
speed of light in vacuum. Thus, the only difference between the two for a given emission
region (particle population) is the magnetic field B, bulk Lorentz factor
$\rm \Gamma$, and the external radiation field $\rm u_{iso, \ast}$ associated with
the other AGN components like BLR, IR torus etc\footnote{Note that there exit other 
radiation fileds like extragalactic Background light (EBL), Cosmic Microwave Background
(CMB). However, their energy densities are much below those from the AGNs components.}.
The external fields, however, hardly vary in comparison to the jet emission on timescales
of days to week.

Further, the emitted power  in 3C 454.3 is dominated by the GeV $\gamma$-rays,
even in the quiescent phase observed by the \emph{Fermi}-LAT \citep[e.g.]
[]{2014PASJ...66...92L}. Thus, the cooling time of the particles emitting
at GeV energies ($\epsilon_\gamma$, peak of the SED and also at the peak of low energy hump) in the
observer's frame can be estimated as \citep[e.g.][]{2014ApJ...796...61K}
\begin{align}\label{eq:cool}
\rm t_{\rm cool} &\simeq (3m_{\rm e} c/4 \sigma_{\rm T} \Gamma^2 u_{iso,\ast,ir})\times \sqrt{(1+z) \epsilon_\ast/\epsilon_{\rm\gamma}} \nonumber \\
	  & \sim 5 \left(\frac{\xi_{\rm ir}}{0.15}\right)^{-1}\left(\frac{\Gamma}{20}\right)^{-2} 
	  \left(\frac{T_\ast}{1200 K}\right)^{-7/2}\left(\frac{\epsilon_{\rm\gamma}}{1 ~GeV}\right)^{-1/2} \text{min} \nonumber
\end{align}
assuming a blackbody torus field ($\rm u_{iso,\ast,ir}/\epsilon_\ast$) of $\rm T_\ast = 1200 K$ with a covering fraction
of $\rm \xi_{ir} = L_{ir}/L_{disk}$ as observed in PKS 1222+216 \citep{2011ApJ...732..116M}
and a bulk Lorentz factor of 20 \citep{2013AJ....146..120L}. The addition of BLR field will
only lower the cooling timescale further.

The estimated cooling timescale is, thus, too small compared to the duration of
any of the flares as well as the binning duration of the LAT data in the analysis.
Hence, one expects a nearly simultaneous variation at IR/optical and $\gamma$-rays
with profile of light curves governed by the size of the emission region. This
is true for the \emph{SEGMENT 1-3} but not for \emph{SEGMENT 4}. The appearance of
flares in $\gamma$-ray without any counterparts in synchrotron bands (IR/optical)
can be explained as a result of the orientation of magnetic field (Manasvita Joshi
*private communication*). Even
in multi-zone leptonic models that have been advocated for such flares 
\citep{2014ApJ...797..137C}, the problem remains. Thus, the appearance of IR/optical
flare before $\gamma$-ray suggest some dynamical effect associated with the magnetic
field, bulk motion, and/or the external radiation field. Since IR/optical emission, for
a given particle distribution, depends only on magnetic field; the IR/optical flare
without $\gamma$-ray could be just due to change in orientation and/or magnitude
of the magnetic field \citep{2013ApJ...763L..11C}. This interpretation is also supported
by the X-ray \citep[being synchrotron self-Compton origin, e.g.][]{2014PASJ...66...92L}
emission which show the trend observed in the IR/optical. Alternatively, it can be a
combined effect
of changes in magnetic field and a lower bulk Lorentz factor, suggesting the 
origin close to the black hole \citep{2013ApJ...763L..11C}. However, we disfavor
this option as the $\gamma$-ray starts rising while the IR/optical flare is still
in declining mode, demanding a substantial change in Lorentz factor. Moreover, a
decelerating jet model is expected to have asymmetric light curves, contrary
to the symmetric profile of the flares here \citep[e.g.][]{2014MNRAS.442..131K}. Another possible
explanation could be a steeper decline of external radiation field energy density
relative to the magnetic field energy density in the jet \citep{2012ApJ...754..114H,
2012ApJ...760..129J}. This present an interesting picture where the location
is probably at the boundary of transition from BLR to IR dominance as advocated
by multi-wavelength studies \citep{2014MNRAS.441.1899F}. Thus, one can estimate
the typical magnetic field required assuming the equivalence between the external
and magnetic energy density as
\begin{align}
 B &= 4\Gamma T^2 \sqrt{\frac{2\pi \sigma_B}{c}} \nonumber \\
   &= 0.4 \left(\frac{\Gamma}{20}\right) \left(\frac{T}{1200}\right)^2 G
\end{align}
consistent with the typical values required by SED modeling ($\sigma_B$ is
the Stefan-Boltzmann constant).

The strong correlation found here is consistent with other similar studies on the
source with lags of zero to few days \citep{2009ApJ...697L..81B, 2015PASP..127....1L,
2015arXiv150203610T, 2012ApJ...758...72W}. The remarkably similar results during various
$\gamma$-ray flaring states of different amplitude and durations in combination with
a very small cooling timescales, implies an extended $\gamma$-ray emitting region
with almost similar physical conditions with the duration and profile being governed
by the size and number density of the emitting particles. This, in turn, also implies
almost similar spectral evolution during these flaring states. The hardening/softening
of $\gamma$-ray spectra \citep{2015arXiv151102280B} can be understood as a result of
competitive contributions between the BLR and IR torus photons, most likely associated
with the variation in the bulk Lorentz factor. These results are also in agreement
with the correlation studies on FSRQs which show lags between 0 to few 10s of days
\citep{2014ApJ...786..157A, 2012ApJ...756...13B, 2012ApJ...754..114H, 2015ApJ...807...79H}
with a tendency of $\gamma$-ray leading the optical \citep{2012AJ....143...23G,
2014ApJ...797..137C}. Here in this case, however, the case is opposite with IR-optical
emission leading the $\gamma$-rays \citep[but see][]{2012AJ....143...23G} and has also
seen in other blazars as well \citep{2013ApJ...763L..11C}.

\section{Conclusions}\label{sec:conclude}
We performed a correlation analysis of multi-wavelength emission from 3C 454.3
during a high $\gamma$-ray activity period from May 13 - December 24, 2014, which is also
noticed in other energy bands. The study performed over an almost continuous segment 
of data shows a highly correlated variation, almost simultaneous across the
electromagnetic spectrum supporting a co-spatial 
emission, thereby strongly favoring leptonic origin scenarios. Interestingly,
the correlation during this period changes from no lags between IR/optical-$\gamma$-ray
to a lag of $\sim$ 3 days with IR/optical leading the $\gamma$-ray suggesting
a change in magnetic field configuration/strength and/or a declining external field 
as the likely process driving the emission.The similarity of results with previous
studies also suggests that the physical conditions remain more or less similar
during different flaring events with amplitude and durations of flares
being mainly governed by the size and particle/magnetic-energy density respectively.

PK's work at University of Sao Paulo (IAG-USP) is supported by the FAPESP Grant No.
2015/13933-0. Most of the work was done while PK was in IUCAA. ACG's work is
partially supported by Chinese Academy of Sciences (CAS)
President's International Fellowship Initiative (PIFI) Grant No. 2016VMB073.
This research has made use of data, software and web tools of High Energy Astrophysics
Science Archive Research Center (HEASARC), maintained by NASA's Goddard Space Flight
Center and an up-to-date SMARTS optical/near-infrared light curves available at
www.astro.yale.edu/smarts/glast/home.php.

\end{document}